\documentclass[twoside,12pt]{article}

\usepackage{epsfig}

\def\NPA{{\em Nucl. Phys.} {\bf A}}

\def\PLB{{\em Phys. Lett.} {\bf B}}

\def\PRL{{\em Phys. Rev. Lett.} }

\def\PRC{{\em Phys. Rev.} {\bf C}}

\newcommand{\be}{\begin{equation}}
\newcommand{\ee}{\end{equation}}
\newcommand{\bea}{\begin{eqnarray}}
\newcommand{\eea}{\end{eqnarray}}

\begin{document}
\title{ \vspace{1cm} In-medium Properties of Hadrons -- Observables}
\author{T.\ Falter, J.\ Lehr, U.\ Mosel, P.\ Muehlich, M.\ Post
\\
Institut fuer Theoretische Physik, Universitaet Giessen\\
D-35392 Giessen, Germany}
\maketitle
\begin{abstract}
We first briefly review the theoretical basis for calculations of
changes of hadronic properties in dense nuclear matter. These
changes have usually been investigated by means of relativistic
heavy-ion reactions. Here we discuss that observable consequences
of such changes can also be seen in more elementary reactions on
nuclei. Particular emphasis is put on a discussion of actual
observables in photonuclear reactions; we discuss in detail
$\eta$- and vector-meson production. We show that photoproduction
of $\eta$'s can yield essential information on in-medium
properties of the $S_{11}(1535)$ resonance while the $\phi$ meson
properties will probably not be accessible through the $K^+K^-$
decay channel. However, for $\omega$ mesons the $\pi^0\gamma$
decay channel, due to its reduced final state interaction, looks
more promising in this respect. Completely free of final state
interactions is dilepton production in the few GeV range. We show
that the sensitivity of this decay channel to changes of hadronic
properties in medium in photonuclear reactions on nuclei is as
large as in ultrarelativistic heavy ion collisions. Finally we
discuss that hadron production in nuclei at 10 -- 20 GeV photon
energies can give important information on the hadronization
process.

\end{abstract}

%\eject
%\tableofcontents

\section{Introduction}

That hadrons can change their properties and couplings in the
nuclear medium has been well known to nuclear physicists since the
days of the Delta-hole model that dealt with the changes of the
properties of the pion and Delta-resonance inside nuclei
\cite{Ericsson-Weise}. Due to the predominant $p$-wave interaction
of pions with nucleons one observes here a lowering of the pion
branch for small but finite pion momenta, which increases with the
nucleon-density. More recently, experiments at the FSR at GSI have
shown that also the pion rest mass in medium differs from its
value in vacuum \cite{Kienle}. This is interesting since there are
also recent experiments \cite{TAPSsigma} that look for the
in-medium changes of the $\sigma$ meson, the chiral partner of the
pion. Any comparison of scalar and pseudoscalar strength could
thus give information about the degree of chiral symmetry
restoration in nuclear matter.

In addition, experiments for charged kaon production at GSI
\cite{KAOS} have given some evidence for the theoretically
predicted lowering of the $K^-$ mass in medium and the (weaker)
rising of the $K^+$ mass. State-of-the-art calculations of the
in-medium properties of kaons have shown that the usual
quasi-particle approximation for these particles is no longer
justified inside nuclear matter where they acquire a broad
spectral function \cite{Lutz,Tolos}.

At higher energies, at the CERN SPS and most recently at the
Brookhaven RHIC, in-medium changes of vector mesons have found
increased interest, mainly because these mesons couple strongly to
the photon so that electromagnetic signals could yield information
about properties of hadrons deeply embedded into nuclear matter.
Indeed, the CERES experiment \cite{CERES} has found a considerable
excess of dileptons in an invariant mass range from $\approx 300$
MeV to $\approx 700$ MeV as compared to expectations based on the
assumption of freely radiating mesons. This result has found an
explanation in terms of a shift of the $\rho$ meson spectral
function down to lower masses, as expected from theory (see, e.g.,
\cite{Peters,Post,Postneu,Wambach}). However, the actual reason
for the observed dilepton excess is far from clear. Both models
that just shift the pole mass of the vector meson as well as those
that also modify the spectral shape have successfully explained
the data \cite{Cassingdil,RappWam,Rapp}; in addition, even a
calculation that just used the free radiation rates with their --
often quite large -- experimental uncertainties was compatible
with the observations \cite{Koch}. There are also calculations
that attribute the observed effect to radiation from a quark-gluon
plasma \cite{Renk}. While all these quite different model
calculations tend to explain the data, though often with some
model assumptions, their theoretical input is sufficiently
different as to make the inverse conclusion that the data prove
one or another of these scenarios impossible.

One of the authors has, therefore, already some years ago proposed
to look for the theoretically predicted changes of vector meson
properties inside the nuclear medium in reactions on normal nuclei
with more microscopic probes \cite{Erice98,Hirschegg}. Of course,
the nuclear density felt by the vector mesons in such experiments
lies much below the equilibrium density of nuclear matter,
$\rho_0$, so that naively any density-dependent effects are
expected to be much smaller than in heavy-ion reactions.

On the other hand, there is a big advantage to these experiments:
they proceed with the spectator matter being close to its
equilibrium state. This is essential because all theoretical
predictions of in-medium properties of hadrons are based on an
model in which the hadron (vector meson) under
investigation is embedded in cold nuclear matter in equilibrium
and with infinite extension. However, a relativistic heavy-ion
reaction proceeds -- at least initially -- far from equilibrium.
Even if equilibrium is reached in a heavy-ion collision this state
changes by cooling through expansion and particle emission and any
observed signal is built up by integrating over the emissions from
all these different stages of the reaction.

Another in-medium effect arises when particles are produced by
high-energy projectiles inside a nuclear medium. A major experimental
effort at RHIC experiments has gone into the observation of jets in
ultra-relativistic heavy-ion collisions and the determination of
their interaction with the surrounding quark or hadronic matter \cite{Jet}.
A complementary process is given by the latest HERMES results at HERA
for high-energy electroproduction of hadrons off nuclei \cite{Hermes}.
Again, the advantage of the latter experiment is that the nuclear matter
with which the interactions happen is at rest and in equilibrium.

In this lecture note we summarize results that we have obtained in
studies of observable consequences of in-medium changes of
hadronic spectral functions as well as hadron formation in
reactions of elementary probes with nuclei. We demonstrate that
the expected in-medium sensitivity in such reactions is as high as
that in relativistic heavy-ion collisions and that in particular
photonuclear reactions present an independent, cleaner testing
ground for assumptions made in analyzing heavy-ion reactions.

\section{Theory}
A large part of the current interest in in-medium properties of
hadrons comes from the hope to learn something about quarks in
nuclei. Indeed, a very simple estimate shows that the chiral
condensate in the nuclear medium is in lowest order in density
given by \cite{Wambach}
\begin{equation}     \label{qbarq}
\langle \bar{q} q \rangle_{\rm med}(\rho,T) \approx \left( 1 -
\sum_h \frac{\Sigma_h \rho^s_h(\rho,T)}{f_\pi^2 m_\pi^2} \right)
\langle \bar{q} q \rangle_{\rm vac} ~.
\end{equation}
Here $\rho_s$ is the \emph{scalar} density of the hadron $h$ in
the nuclear system and $\Sigma_h$ the so-called sigma-commutator
that contains information on the chiral properties of $h$. The sum
runs over all hadronic states. While (\ref{qbarq}) is nearly
exact, its actual applicability is limited because neither the
sigma-commutators of the higher lying hadrons nor their scalar
densities are known. However, at very low temperatures close to
the groundstate these are accessible. Here $\rho_s \approx \rho_v
\frac{m}{E}$ so that the condensate drops linearly with the
nuclear (vector) density. This drop can be understood in physical
terms: with increasing density the hadrons with a chirally
symmetric phase in their interior fill in more and more space in
the vacuum with its spontaneously broken chiral symmetry. Note
that this is a pure volume effect; it is there already for a free,
non-interacting hadron gas.

How this drop of the scalar condensate translates into observable
hadron masses is not uniquely prescribed. The only rigorous
connection is given by the QCD sum rules that relates an integral
over the hadronic spectral function to a sum over combinations of
quark- and gluon-condensates with powers of $1/Q^2$. The starting
point for this connection is the current-current correlator
\begin{equation}
\Pi_{\mu\nu}(q) = i \int\!\! d^4\!x \, e^{iqx} \langle T j_\mu(x)
j_\nu(0) \rangle    \, ,
  \label{eq:curcur}
\end{equation}
where the current is expressed in terms of quark-field operators.
This correlator can be decomposed \cite{Leupold}
\begin{equation}
  \label{eq:decompmunu}
\Pi_{\mu\nu}(q) = q_\mu q_\nu R(q^2) - g_{\mu\nu} \Pi^{\rm
isotr}(q^2)    \,.
\end{equation}
The QCD sum rule then reads
\begin{eqnarray}
R^{{\rm OPE}}(Q^2) & = & {\tilde c_1 \over Q^2} + \tilde c_2 -{Q^2
\over \pi} \int\limits_0^\infty \!\! ds \, {{\rm Im}R^{{\rm
HAD}}(s) \over (s+Q^2)s}
  \label{eq:opehadr}
\end{eqnarray}
with $Q^2:= -q^2 \gg 0$ and some subtraction constants $\tilde
c_i$. Here $R^{\rm OPE}$ represents a Wilson's operator expansion
of the current-current correlator in terms of quark and gluon
degrees of freedom in the space-like region. On the other hand,
$R^{{\rm HAD}}(s)$ in (\ref{eq:opehadr}) is the same object for
time-like momenta, represented by a parametrization in terms of
hadronic variables. The dispersion integral connects time- and
space-like momenta. Eq. (\ref{eq:opehadr}) also connects the
hadronic with the quark world. It allows -- after the technical
step of a Borel transformation -- to determine parameters in a
hadronic parametrization of $R^{{\rm HAD}}(s)$ by comparing the
lhs of this equation with its rhs.

The operator product expansion of $R^{\rm OPE}$ on the lhs
involves quark- and gluon condensates \cite{Leupold,LeupoldMosel};
of these only the two-quark condensates are reasonably well known,
whereas our knowledge about already the four-quark condensates is
rather sketchy.

Leupold et al. \cite{Leupold,LeupoldMosel} have shown that using
a realistic parametrization of the hadronic spectral function on
the rhs of (\ref{eq:opehadr})
\begin{eqnarray}
{\rm Im}R^{{\rm HAD}}(s) & = & \Theta(s_0 -s) \, {\rm Im}R^{{\rm
RES}}(s) + \Theta(s -s_0) \, {1 \over 8 \pi} \left( 1+ {\alpha_s
\over \pi} \right)
  \label{eq:pihadans}
\end{eqnarray}
where $s_0$ denotes the threshold between the low energy region
described by a spectral function for the lowest lying resonance,
${\rm Im}R^{{\rm RES}}$, and the high energy region described by a
continuum calculated from perturbative QCD. The second term on the
rhs of (\ref{eq:pihadans}) represents the QCD perturbative result
that survives when $s \to \infty$. Using the measured, known
vacuum spectral function for $R^{\rm HAD}$ allows one to obtain
information about the condensates appearing on the lhs of
(\ref{eq:opehadr}). On the other hand, modelling the
density-dependence of the condensates yields information on the
change of the hadronic spectral function when the hadron is
embedded in nuclear matter. Since the spectral function appears
under an integral the information obtained is not very precise.
However, Leupold et al. have shown \cite{Leupold,LeupoldMosel}
that the QCDSR provides important constraints for the hadronic
spectral functions in medium, but it does not fix them. Recently
Kaempfer et al have turned this argument around by pointing out
that measuring an in-medium spectral function of the $\omega$
meson could help to determine the density dependence of the
higher-order condensates \cite{Kaempfer}.

Thus models are needed for the hadronic interactions. The
quantitatively reliable ones can at present be based only on
'classical' hadrons and their interactions. Indeed, in lowest
order in the density the mass and width of an interacting hadron
in nuclear matter at zero temperature and vector density $\rho_v$
are given by (for a meson, for example)
\begin{eqnarray}     \label{trho}
{m^*}^2 = m^2 - 4 \pi \Re f_{m N}(q_0,\theta = 0)\, \rho_v
\nonumber \\
m^* \Gamma^* = m \Gamma^0 -  4 \pi \Im f_{mN}(q_0,\theta = 0)\,
\rho_v ~.
\end{eqnarray}
Here $f_{mN}(q_0,\theta = 0)$ is the forward scattering amplitude
for a meson with energy $q_0$ on a nucleon. The width $\Gamma^0$
denotes the free decay width of the particle. For the imaginary
part this is nothing other than the classical relation $\Gamma^* -
\Gamma^0 = v \sigma \rho_v$ for the collision width, where
$\sigma$ is the total cross section. This can easily be seen by
using the optical theorem.

Note that such a picture also encompasses the change of the chiral
condensate in (\ref{qbarq}), obtained there for non-interacting
hadrons. If the spectral function of a non-interacting hadron
changes as a function of density, then in a classical hadronic
theory, which works with fixed (free) hadron masses, this change
will show up as an energy-dependent interaction and is thus
contained in any empirical phenomenological cross section.

\begin{figure}[tb]

\begin{center}

\begin{minipage}[t]{8 cm}
\epsfig{file=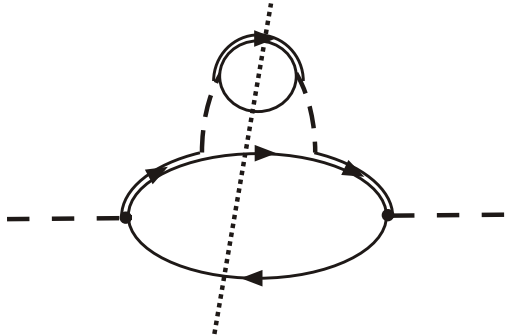,scale=1}
\end{minipage}
\begin{minipage}[t]{12cm}
\caption{Selfenergy diagram for a meson coupling to an in-medium
resonance. The dotted line shows a possible cut through the
diagram to generate the imaginary part of the selfenergy (from
\cite{Postneu}). Here a nucleon is represented by a solid line, a meson by
a dashed line and a baryon resonance by a double line. \label{inmedself}}
\end{minipage}

\end{center}

\end{figure}

Unfortunately it is not a-priori known up to which
densities the low-density expansion (\ref{trho}) is useful.
Post et al. have recently investigated this question in a
coupled-channel calculation of selfenergies \cite{Postneu}.
Their analysis comprises pions, $\eta$-mesons and $\rho$-mesons as
well as all baryon resonances with a sizeable coupling to any of
these mesons. Fig.\ \ref{inmedself} shows a typical selfenergy graph
in their calculation. Of particular importance are the
$\Delta(1232)$, the $S_{11}(1535)$ and the $D_{13}(1520)$ resonances
for the pion, the $\eta$-meson and the $\rho$-meson, respectively.

\begin{figure}[tb]

\begin{center}

\begin{minipage}[t]{12 cm}
\centerline{\epsfig{file=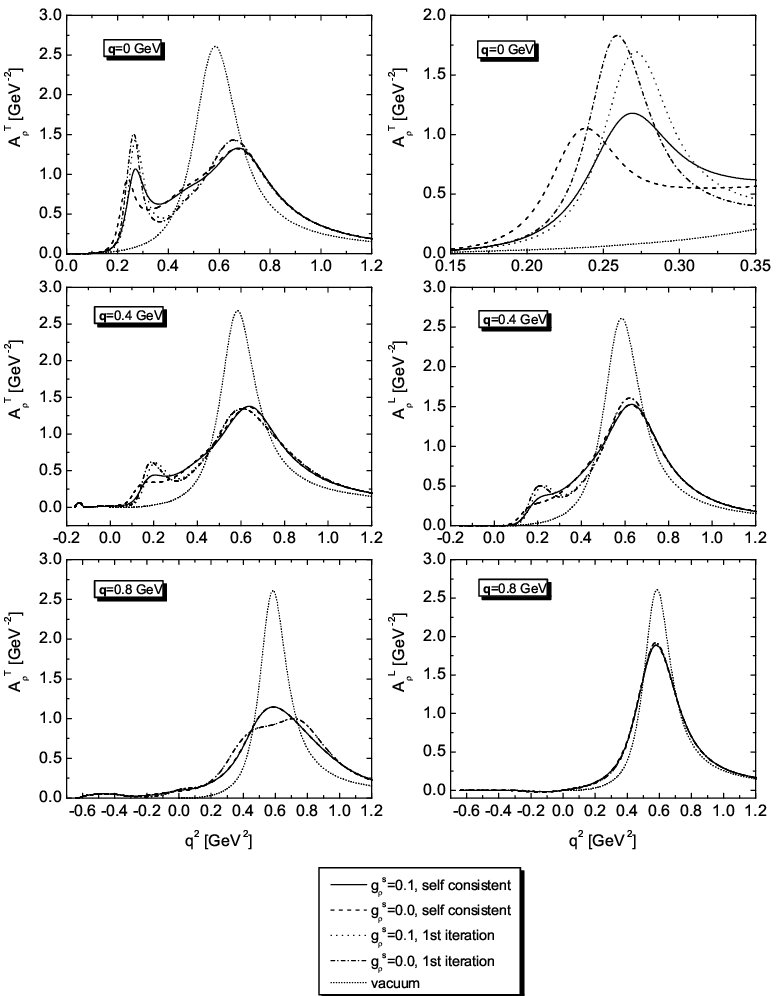,scale=1.4}}
\end{minipage}
\begin{minipage}[t]{12 cm}
\caption{Spectral function of the $\rho$-meson at normal nuclear
matter density for various momenta. Shown are the transverse and
longitudinal spectral functions which are degenerate at $\vec{q} =
0$. Also shown are the effects of selfconsistency (from
\cite{Postneu}). \label{rhospectral}}
\end{minipage}

\end{center}

\end{figure}

The authors of \cite{Postneu} find that already for densities less
than $0.5 \rho_0$ the linear scaling of the selfenergies inherent
in (\ref{trho}) is badly violated for the $\rho$ and the $\pi$
mesons, whereas it is a reasonable approximation for the $\eta$
meson, but even in the latter case already at $0.5 \rho_0$ the
selfenergy does not scale exactly with $\rho$. Reasons for this
deviation from linearity are Fermi-motion, Pauli-blocking,
selfconsistency and short-range correlations. For different mesons
different sources of the discrepancy prevail: for the $\rho$ and
$\eta$ mesons the iterations act against the low-density theorem
by inducing a strong broadening for the $D_{13}(1520)$ and a
slightly repulsive mass shift for the $S_{11}(1535)$ nucleon
resonances to which the $\rho$ and the $\eta$ meson, respectively,
couple. The effects of self-consistency on the spectral function
of the $\rho$-meson can nicely be seen in Fig. \ref{rhospectral}.
The figure shows that the spectral function of a $\rho$  meson at
rest in the nuclear medium (upper left, solid curve) exhibits a
remarkable structure that is generated by the strong coupling to
the $D_{13}(1520)$ resonance. While this structure disappears at
the highest momentum considered here, the spectral function for
the transverse vector meson (lower left) is still considerably
different from that of a $\rho$ in vacuum. Notice also that the
longitudinal spectral function (right column) retains more of the
original vector meson structure.

\section{Particle Production on Nuclei -- Observables}

Traditionally, heavy-ion reactions are assumed to be the reactions
of choice for an investigation of in-medium properties. After all,
here very high nuclear densities ($2 - 10 \rho_0$) can be reached
with present day's accelerators so that any density-dependent
effect would be magnified. However, any explanations of observed
effects always suffer from an inherent inconsistency: The observed
signal, on one hand, necessarily integrates over many different
stages of the collision -- nonequilibrium and equilibrium, the
latter at various densities and temperatures. On the other hand,
the theoretical input is always calculated under the simplifying
assumption of a hadron in nuclear matter in equilibrium. We have
therefore looked for possible effects in reactions that proceed
much closer to equilibrium, i.e. reactions of elementary probes
such as protons, pions, and photons on nuclei. Of course, the
densities probed in such reactions are always $\le \rho_0$, with
most of the nucleons actually being at about $0.5 \rho_0$. On the
other hand, the target is obviously stationary and the reaction
proceeds much closer to (cold) equilibrium than in a relativistic
heavy-ion collision.

It is thus a quantitative question if any observable effects of
in-medium changes of hadronic properties survive if the densities
probed are always $\le \rho_0$. With the aim of answering this
question we have over the last few years undertaken a number of
calculations for proton- \cite{Bratprot}, pion-
\cite{Weidmann,Effepi} and photon- \cite{Effephot} induced
reactions. All of them have one feature in common: they treat the
final state incoherently in a coupled channel transport
calculation that allows for elastic and inelastic scattering of,
particle production by and absorption of the produced hadrons (vector
mesons). A new feature of these calculations is that vector mesons
with their correct spectral functions can actually be produced and
transported consistently. This is quite an advantage over earlier
treatments \cite{Brat-Cass} in which the mesons were always
produced and transported with their pole mass and their spectral
function was later on folded in only for their decay.

In all photonuclear calculations above energies of about 1 GeV we
account for coherent initial state interactions of the incoming
photon which lead to shadowing. In an interaction with a bound
nucleon the shadowed incoming photon then produces a hadronic
final state which cascades through the nucleus
\cite{Effe,FalterShad}. The incoherent final state interactions
are treated as just described by means of a semi-classical
coupled-channel transport theory. For details of the model see
Ref. \cite{Effephot} and references therein.

\subsection{\it $\eta$ Production}
We first look at the prospects of using reactions with hadronic
final states and discuss the photo-production of $\eta$-mesons on
nuclei as a first example. These mesons are unique in that they
are sensitive to one dominating resonance only (the
$S_{11}(1535)$) so that one may hope to learn something about the
properties of this resonance inside nuclei. Experiments for this
reaction were performed both by the TAPS collaboration \cite{TAPS}
and -- for higher energies -- at KEK \cite{KEK}.

Estimates of the collisional broadening of the $S_{11}(1535)$
resonance have given a collision width of about 35 MeV at $\rho_0$
\cite{EffeHom}. The more recent, and more refined, selfconsistent
calculations of \cite{Postneu} give a very similar value for this
resonance. In addition, a dispersive calculation of the real part
of the selfenergy for the resonance at rest gives only an
insignificant shift of the resonance position. Thus any
momentum-dependence of the selfenergy as observed in
photon-nucleus data can directly be attributed to binding energy
effects \cite{Lehreta}.

\begin{figure}

\begin{center}

\begin{minipage}[t]{8 cm}
\centerline{\epsfig{file=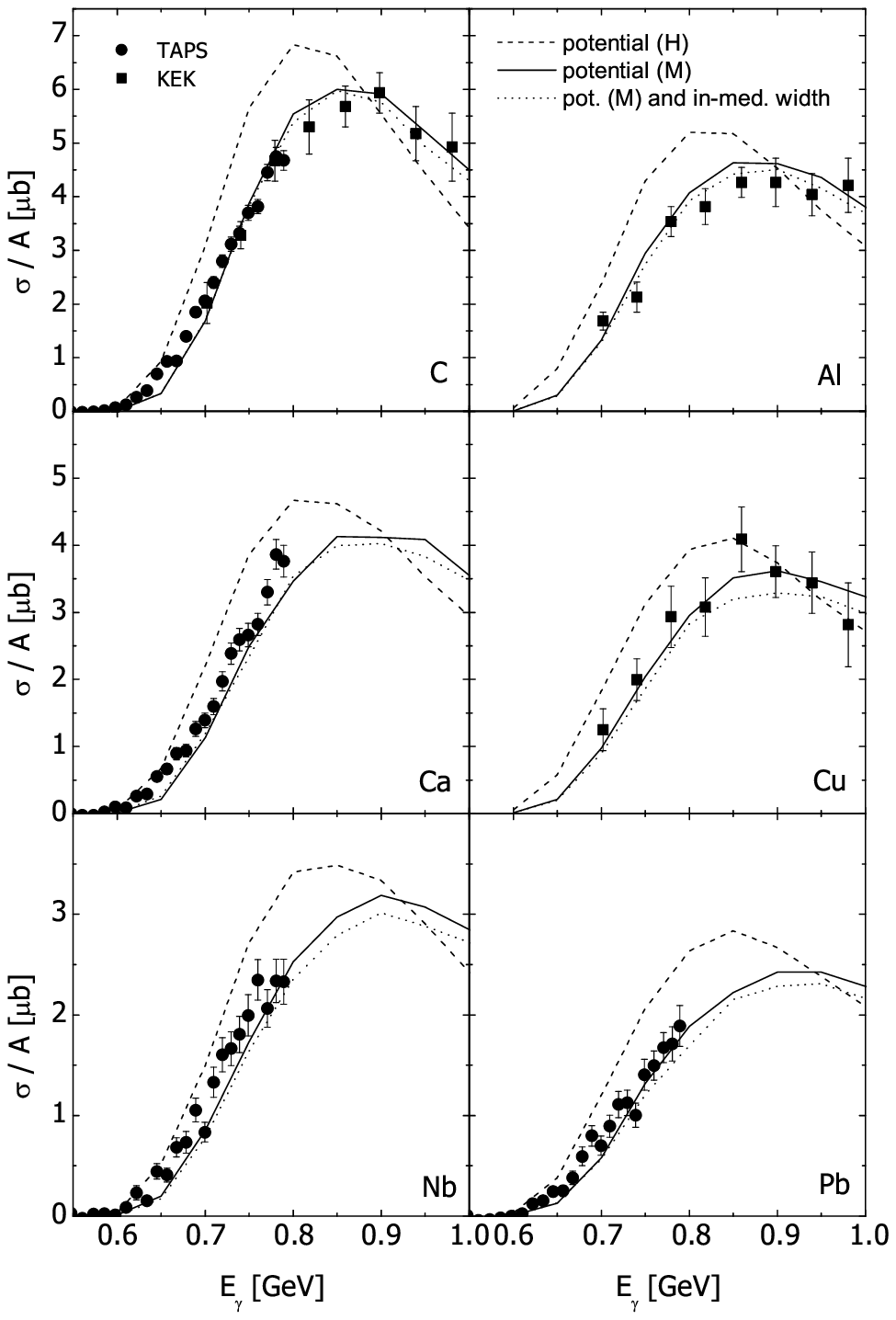,scale=0.8}}
\end{minipage}
\begin{minipage}[t]{12 cm}
\caption{Results for the reaction $\gamma + A \to \eta X$. The
dashed and solid lines correspond to the use of two different
potentials (see text). The data are from \cite{TAPS} (circles) and
\cite{KEK} (squares) (from \cite{Lehreta}). \label{etaphot}}
\end{minipage}

\end{center}

\end{figure}

The results of this calculation \cite{Lehreta} are shown in Fig.
\ref{etaphot}.
The dashed line in Fig. \ref{etaphot} gives the results of a
calculation employing a momentum-independent potential for the
$S_{11}(1535)$ resonance with the same depth at $\rho_0$ as for
the nucleons. It is clearly seen that it rises too steeply and
peaks at a two low energy compared to the data. The solid line,
that describes all available data over a wide energy range very
well, employs a momentum-dependence in the nucleon and resonance
potentials taken from earlier fits to heavy-ion reaction data. For
the present discussion the relevant feature of this potential is
that it vanishes for momenta around 800 MeV. Thus, for photon
energies around this values, close to the peak position in the
data, the resonance is no longer bound whereas the hit nucleon is
bound by about 75 MeV at $\rho_0$. Thus, effectively the resonance
is shifted upwards by this amount; this is just what the data
show. In addition, Fig. \ref{etaphot} also contains a dotted line,
exhibiting the effects of the in-medium width of the
$S_{11}(1535)$ resonance. Since the latter is small, also its
effect on the eta-production cross section is small.

\subsection{\it $\phi$ Production}

As a second example we look into the photoproduction of
$\phi$-mesons in nuclei. Our calculations of this reaction have
been motivated by the idea to learn about the $\phi$ in-medium
properties by measuring the $K^+K^-$ mass distribution
\cite{MuehlPhi}. It has been proposed to perform such an
experiment at the Spring8 facility in Osaka \cite{Osaka}. In
principle the $\phi$ meson forms a unique probe since it is the
only resonance decaying into $K^+K^-$ pairs in the considered mass
region. Theoretical approaches to the $\phi$ in-medium properties
find a broadening of the $\phi$ meson of up to 40 MeV at nuclear
matter density and vanishing $\phi$ momentum \cite{OsetPhi}. While
this value exceeds the vacuum value of 4.4 MeV by about one order
of magnitude, a possible mass shift of the $\phi$ is expected to
be of the order of only $-10$ MeV to $-30$ MeV \cite{Hatsuda} so
that the $\phi$ meson is expected to survive as a proper
quasiparticle even at nuclear matter density $\rho_0$
\cite{WeisePhi}. Thus the expected modification might lead to
observable consequences because of the small width of the $\phi$
meson.

\begin{figure}
\begin{center}
\begin{minipage}[t]{8 cm}
\centerline{\epsfig{file=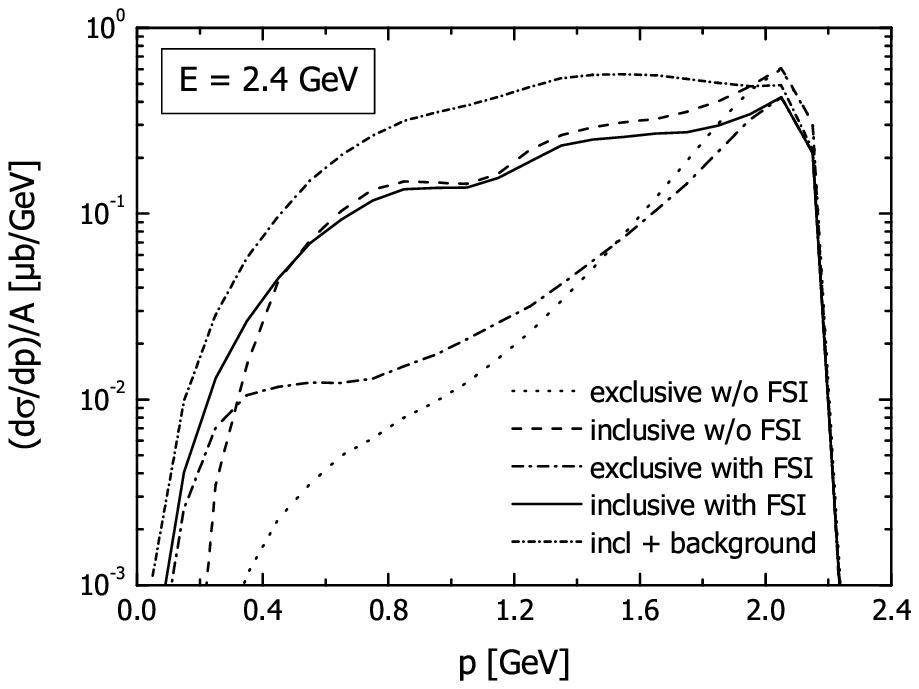,scale=.9}}
\end{minipage}
\begin{minipage}[t]{12 cm}
\caption{Momentum differential cross section for $K^+K^-$
photoproduction off $^{40}$Ca (from
\cite{MuehlPhi}).\label{phi01}}
\end{minipage}
\end{center}
\end{figure}

Only vector mesons decaying inside the nucleus may reveal
informations about their in-medium properties. Hence, the decay
length of these vector mesons $L_V=p_V/(m_V\Gamma_V)$ has to be of
the order of the nuclear radius. In order to produce $\phi$ mesons
with the required low momenta in the target rest frame the
incident photon energy has to be chosen not too close above
threshold ($\sim$ 1.5 GeV). Simulations at higher photon energies
require to include also inclusive photoproduction processes. This
is illustrated in Fig. \ref{phi01}, which shows the momentum
distribution of $\phi$ mesons with (solid line) and without FSI
(dashed) from a nuclear target ($^{40}$Ca). It is clearly seen
that the cross section at small $\phi$ momenta is dominated by the
inclusive production processes. In the simulations inclusive
photoproduction of vector mesons has been accounted for by means
of the hadronic event generator FRITIOF employing the vector meson
dominance model in order to describe the photon-hadron coupling
\cite{MuehlPhi}. Also visible from Fig. \ref{phi01} is the fact,
that the FSI of the $\phi$, through their non-absorptive part,
produce an accumulation of $\phi$ mesons in the low-momentum part
of the spectrum. The number of $\phi$ mesons decaying in the
medium is greatly enhanced by this effect.

\begin{figure}
\begin{center}
\begin{minipage}[t]{10 cm}
\centerline{\epsfig{file=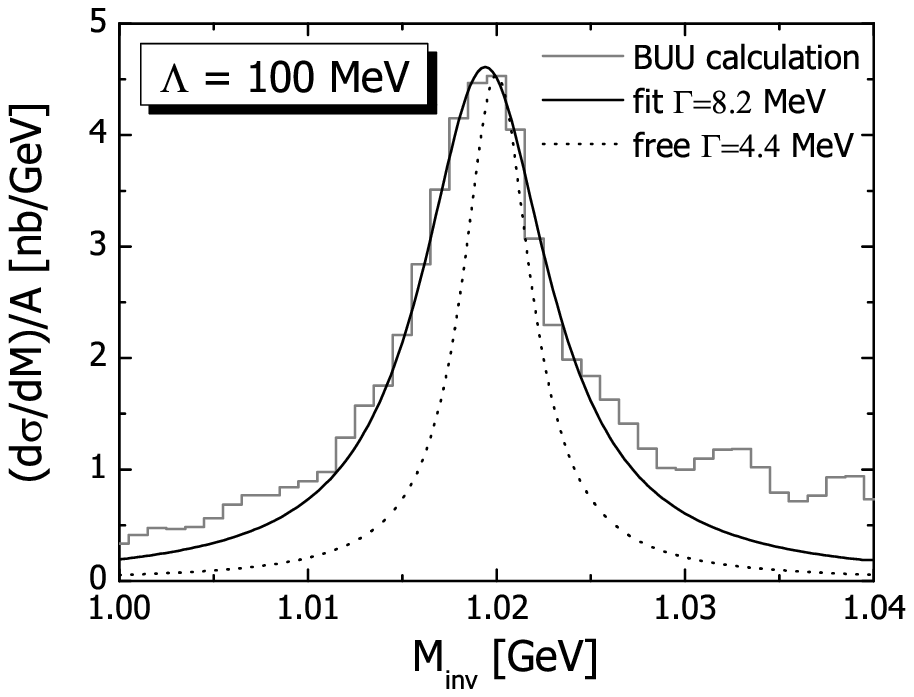,scale=1.0}}
\end{minipage}
\begin{minipage}[t]{12 cm}
\caption{Mass differential cross section for $K^+K^-$
photoproduction off $^{40}$Ca with a momentum cutoff of
$\Lambda=150$ MeV at 2.4 GeV photon energy (from
\cite{MuehlPhi}).\label{phi02}}
\end{minipage}
\end{center}
\end{figure}

Including collisional broadening by means of the low-density
expansion (\ref{trho}) and applying a rather severe momentum
cutoff to the three momentum of the detected $K^+K^-$ pair, we
find an observable broadening of the invariant mass spectrum by
almost a factor of two, see Fig. \ref{phi02}. However, accounting
also for the electromagnetic potential which already in $^{40}$Ca
has a depth of 11 MeV, the average decay density of reconstructed
$\phi$ mesons decreases considerably. As a consequence the
obtained invariant mass spectrum again is very close to the free
one (dotted in Fig. \ref{phi02}). Also the dropping $\phi$ mass
(30 MeV at $\rho_0$ in the simulation) has no remarkable effect,
since as long as the $K^+$ and $K^-$ properties are not changed in
the medium $\phi$ mesons with considerably reduced masses do not
decay into $K^+K^-$ pairs. This also leads to a reduced average
decay density.

\begin{figure}
\begin{center}
\begin{minipage}[t]{8 cm}
\centerline{\epsfig{file=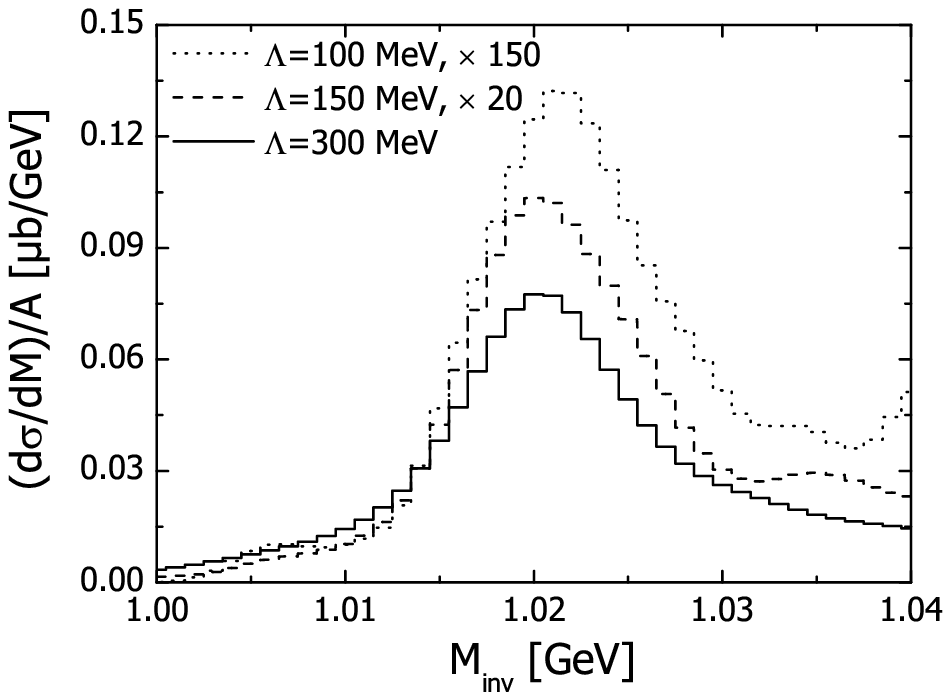,scale=.9}}
\end{minipage}
\begin{minipage}[t]{12 cm}
\caption{Mass differential cross section for $K^+K^-$
photoproduction off $^{40}$Ca at $E_{\gamma}=2.4$ GeV including
density-dependent mass modifications of the kaons, collisional
broadening of the $\phi$ meson and the Coulomb potential (from
\cite{MuehlPhi}).\label{phi03}}
\end{minipage}
\end{center}
\end{figure}

As already mentioned, also the kaon properties receive a rather
dramatic change in the nuclear environment \cite{Schaffner}. Kaons
from $\phi$ decays that propagate through a potential gradient
change their momenta and, hence, also the invariant mass of any
$K^+K^-$ pair. This effect leads to an accumulation of strength on
the high mass side of the invariant mass spectrum, see Fig.
\ref{phi03}. In summary, we find, that due to the medium
modification of the $\phi$ decay products (the kaons) one cannot
learn anything about the $\phi$ properties in the
$A(\gamma,K^+K^-)X$ reaction, but rather about the existence of an
in-medium change of the $K^+$ and $K^-$ mesons.

\subsection{\it $\omega$ Production}

In contrast to the results for the $\phi$ meson, the situation
looks more promising for the $\omega$ meson \cite{MuehlOm}. On the
one hand, the $\omega$ is much lighter than the $\phi$ and can,
therefore, be produced with lower momentum inside nuclei. On the
other hand, the $\pi^0\gamma$ decay mode is not influenced by the
electromagnetic potential and also FSI effects might be less
pronounced because of the weak photon-nucleon coupling. An
experiment measuring the $A(\gamma,\pi^0\gamma')X$ reaction is
presently being analyzed by the TAPS/Crystal Barrel collaborations
at ELSA \cite{Messch}. The varying theoretical predictions for the
$\omega$ mass (640-765 MeV) \cite{Klingl} and width (up to 50 MeV)
\cite{Friman,Weidmann} in nuclear matter at rest encourage the use
of such an exclusive probe to learn about the $\omega$ spectral
distribution in nuclei.

\begin{figure}
\begin{center}
\begin{minipage}[t]{10 cm}
\centerline{\epsfig{file=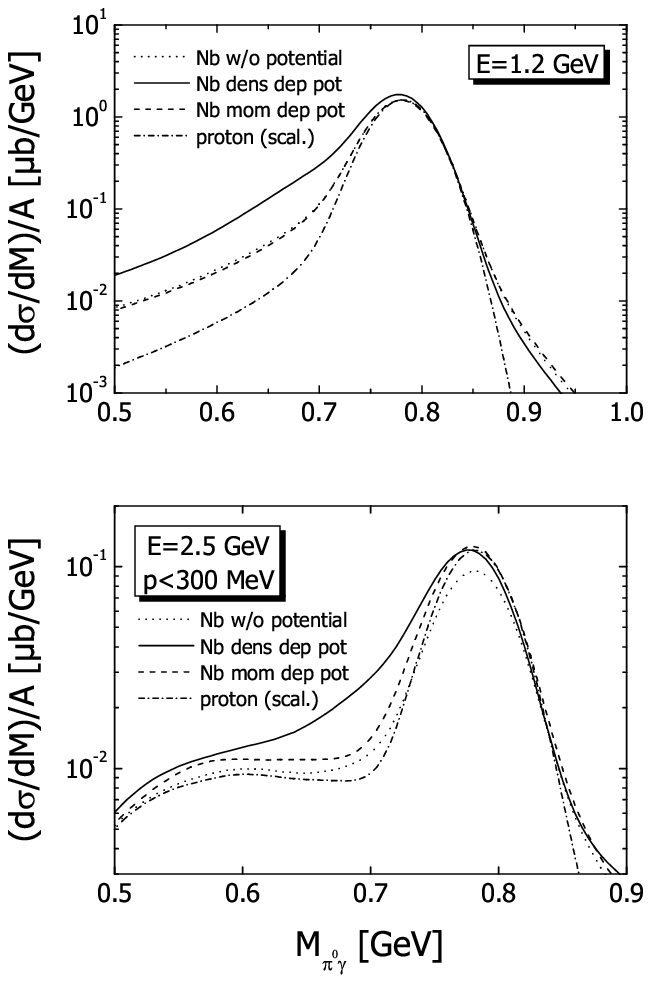,scale=1.4}}
\end{minipage}
\begin{minipage}[t]{12 cm}
\caption{Mass differential cross section for $\pi^0\gamma$
photoproduction off $^{93}$Nb at two different beam energies (from
\cite{MuehlOm}). The results at $E_{\gamma}=2.5$ GeV are obtained
with a $\omega$ momentum cutoff of 300 MeV. The proton cross
section is normalized to the cross section on Niobium obtained
with the momentum dependent potential at the $\omega$ pole mass
$m=0.782$ GeV. \label{omega01}}
\end{minipage}
\end{center}
\end{figure}

Simulations have been performed at 1.2 GeV and 2.5 GeV photon
energy, which cover the accessible energies of the TAPS/Crystal
Barrel experiment. After reducing the combinatorial and
rescattering background by applying kinematic cuts on the outgoing
particles, we obtain rather clear signals for a dropping $\omega$
mass inside nuclei. The solid line in Fig. \ref{omega01} shows
results of a calculation including both collisional broadening and
a lowering of the $\omega$ mass. It exhibits a shoulder on the
low-mass side of the peak and a clear excess there over the proton
distribution, also shown in the figure. This has to be compared to
the dashed line, which includes collisional broadening only.
Therefore, in this particular case it is possible to disentangle
rather trivial in-medium changes (collisional broadening) from
more interesting medium-modifications (dropping mass), which may
be related to a change of the chiral condensate at finite nuclear
density \cite{BrownRho}.

Employing the low-density expansion of Eq. (\ref{trho}) to
determine the effective in-medium mass of the $\omega$ one obtains
a momentum dependence which is driven by the energy dependence of
the $\omega$-nucleon total cross section. Such a simple-minded
approach yields a considerably reduced mass for $\omega$ mesons at
rest and a fast increase of the effective mass for $\omega$ mesons
moving with finite velocity, arriving rather fast at values close
to the $\omega$ vacuum mass \cite{MuehlOm}. The results of
simulations adopting this momentum dependent mass are also shown
in Fig. \ref{omega01} (dash-dotted line). As expected the results
of these calculations are again very similar to the results
obtained without any real $\omega$ potential. Measurements with
varying momentum cuts may help to disentangle the density and
momentum dependence of the in-medium self-energy of the $\omega$
meson.

\subsection{\it Dilepton Production}

\begin{figure}

\begin{center}

\begin{minipage}[t]{8 cm}
\centerline{\epsfig{file=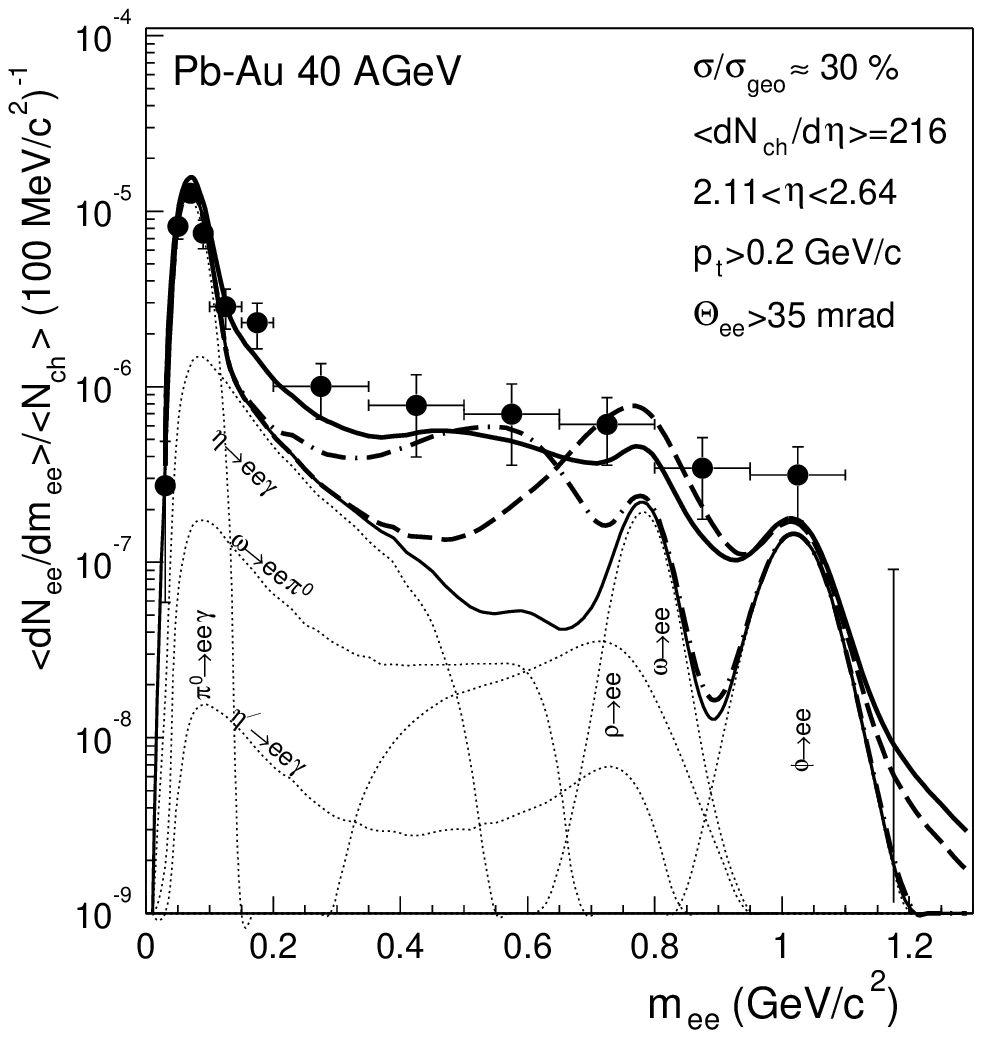,scale=0.8}}
\end{minipage}
\begin{minipage}[t]{12 cm}
\caption{Invariant  dilepton mass spectrum
 obtained with the CERES experiment in Pb + Au collisions at 40
 AGeV (from \cite{CERES}). The thin curves give the contributions
 of individual hadronic sources to the total dilepton yield, the
 fat solid (modified spectral function) and the dash-dotted
 (dropping mass only) curves give the results of calculations
 \cite{Rapp} employing an in-medium modified spectral function of the vector
 mesons.} \label{CERES}
\end{minipage}

\end{center}

\end{figure}

Dileptons, i.e. electron-positron pairs, in the outgoing channel
are an ideal probe for in-medium properties of hadrons since they
-- in contrast to hadronic probes -- experience no strong final
state interaction. A first experiment to look for these dileptons
in heavy-ion reactions was the DLS experiment at the BEVALAC in
Berkeley \cite{DLS}. Later on, and in a higher energy regime, the
CERES experiment has received a lot of attention for its
observation of an excess of dileptons with invariant masses below
those of the lightest vector mesons \cite{CERES}. Explanations of
this excess have focused on a change of in-medium properties of
these vector mesons in dense nuclear matter (see e.g.\
\cite{Cassingdil,RappWam}). The radiating sources can be nicely
seen in Fig.~\ref{CERES} that shows the dilepton spectrum obtained
in a low-energy run at 40 AGeV together with the elementary
sources of dilepton radiation.

\begin{figure}

\begin{center}

\begin{minipage}[t]{8 cm}
\centerline{\epsfig{file=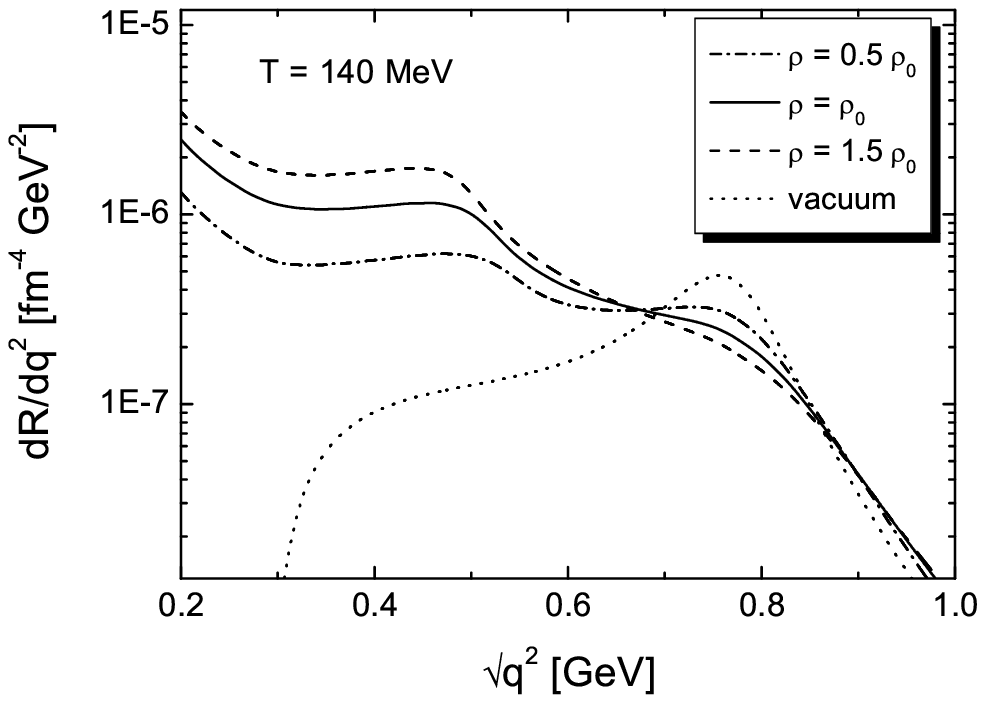,scale=0.8}}
\end{minipage}
\begin{minipage}[t]{12 cm}
\caption{Momentum integrated dilepton rates at a temperature $T =
140$ MeV for three densities (from \cite{Postneu}).} \label{Tdilept}
\end{minipage}

\end{center}

\end{figure}

The figure exhibits clearly the rather strong contributions of the
vector mesons -- both direct and through their Dalitz decay -- at
invariant masses above about 500 MeV. If this strength is shifted
downward, caused by an in-medium change of the vector-meson
spectral functions, then the observed excess can be explained as
has been shown by various authors (see e.g.\ \cite{Brat-Cass} for
a review of such calculations). The results of a new calculation,
based on the spectral function in Fig. \ref{rhospectral}, are shown
in Fig. \ref{Tdilept}.
The strong amplification of the dilepton rate at small invariant
masses is caused by the photon propagator, which contributes $\sim
1/q^4$ to the cross section. Any strength in the spectral function
at finite density at small $q^2$ is thus dramatically amplified.

In view of the uncertainties in interpreting these results
discussed earlier we have studied the dilepton production in
reactions on nuclear targets involving more elementary
projectiles. It is not \emph{a priori} hopeless to look for
in-medium effects on dilepton production in ordinary nuclei: Even
in relativistic heavy-ion reactions that reach baryonic densities
of the order of 3 - 10 $\rho_0$ many observed dileptons actually
stem from densities that are much lower than these high peak
densities. Transport simulations have shown \cite{Brat-Cass} that
even at the CERES energies about 1/2 of all dileptons come from
densities lower than $2 \rho_0$. This is so because in such
reactions the pion-density gets quite large in particular in the
late stages of the collision, where the baryonic matter expands
and its density becomes low again. Correspondingly many vector
mesons are formed (through $\pi + \pi \to \rho$) late in the
collision and their decay to dileptons thus happens at low baryon
densities.

\begin{figure}

\begin{center}

\begin{minipage}[t]{7 cm}
\centerline{\epsfig{file=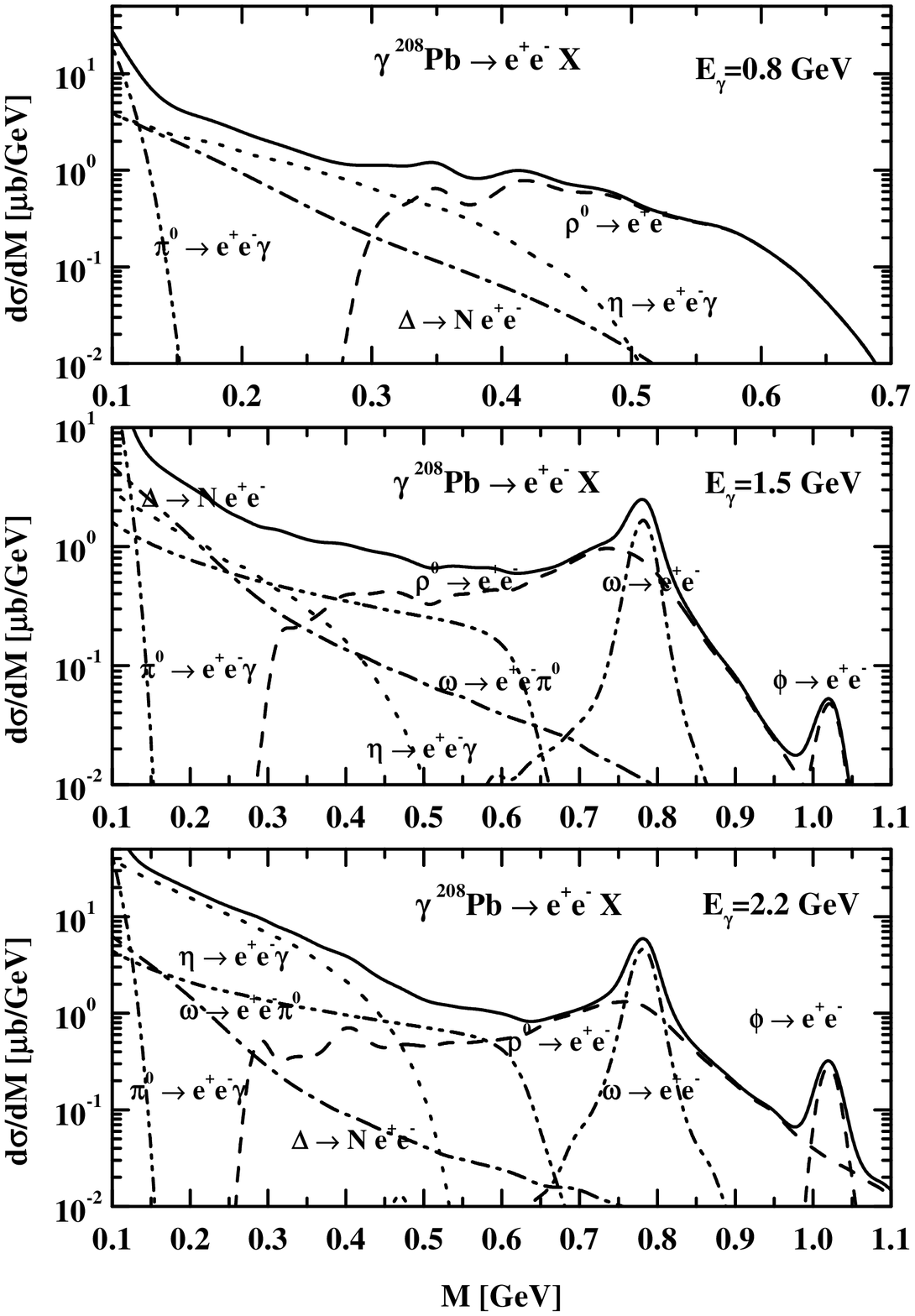,scale=0.7}}
\end{minipage}
\begin{minipage}[t]{12 cm}
\caption{Hadronic contributions to dilepton invariant mass spectra
for $\gamma + ^{208}Pb$ at the three photon energies given (from
\cite{Effephot}). Compare with Fig.~\ref{CERES}.} \label{Fige+e-}
\end{minipage}

\end{center}

\end{figure}

\begin{figure}

\begin{center}

\begin{minipage}[t]{8 cm}
\centerline{\epsfig{file=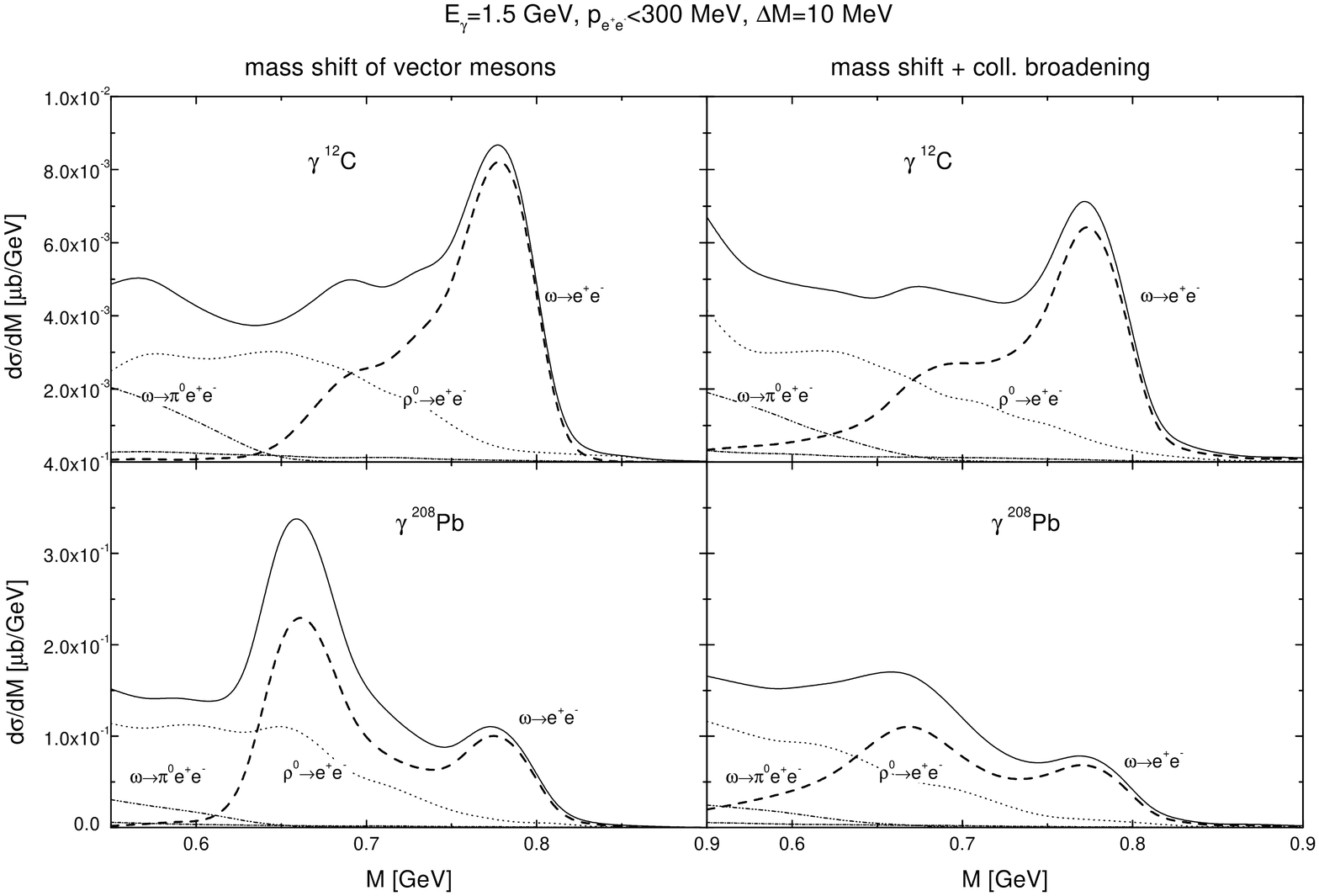,scale=0.55}}
\end{minipage}
\begin{minipage}[t]{12 cm}
\caption{Dilepton mass yield with a dilepton-momentum cut of 300
MeV. Shown on the left are results of a calculation that uses only
a shift of the pole mass of the vector mesons. On the right,
results are given for a calculation using both mass shift and
collisional broadening (from \cite{Effe}).} \label{Figdlim}
\end{minipage}

\end{center}

\end{figure}

A typical result of such a calculation for the dilepton yield --
after removing the Bethe-Heitler component -- is given in Fig.
\ref{Fige+e-}.
Comparing this figure with Fig. \ref{CERES} shows that in a
photon-induced reaction at 1 - 2 GeV photon energy exactly the
same sources, and none less, contribute to the dilepton yield as
in relativistic heavy-ion collisions at 40 AGeV! The question now
remains if we can expect any observable effect of possible
in-medium changes of the vector meson spectral functions in medium
in such an experiment on the nucleus where -- due to surface
effects -- the average nucleon density is below $\rho_0$. This
question is answered, for example, by the results of Fig.\
\ref{Figdlim}.
This figure shows the dilepton spectra to be expected if a
suitable cut on the dilepton momenta is imposed; with this cut
slow vector mesons are enriched. In the realistic case shown on
the right, which contains both a collision broadening and a mass
shift, it is obvious that a major signal is to be expected: in the
heavy nucleus $Pb$ the $\omega$-peak has completely disappeared
from the spectrum. The sensitivity of such reactions is thus as
large as that observed in ultrarelativistic heavy-ion reactions.

An experimental verification of this prediction would be a major
step forward in our understanding of in-medium changes\footnote{An
experiment at JLAB is under way \cite{Weygand}.}. It would
obviously present a purely hadronic base-line to all data on top
of which all 'fancier' explanations of the CERES effect in terms
of radiation from a QGP and the such would have to live.

\section{Hadron Formation}
High energy photo- and electroproduction of hadrons on complex
nuclei offers a promising tool to study the physics of hadron
formation. The interaction of the (virtual) photon with a nucleon
leads to the production of several high energy particles. Due to
the soft momentum scales involved in the hadronization process the
formation of the final hadrons is of nonperturbative nature and
not well understood. A simple estimate for the hadron formation
time can be given by the time that the constituents need to travel
a hadronic radius. It should therefore be of the order of 0.5--0.8
fm/c in the rest frame of the hadron. Due to time dilatation the
formation lengths of high energetic particles can reach distances
in the lab frame that are of the order of the nuclear radius. Thus
nuclei can serve as a kind of 'microdetector' for the
determination of formation times as well as the interactions while
the final hadron is still being formed. Obviously, the latter is
closely related to color transparency.

The HERMES collaboration \cite{Hermes} has investigated the
multiplicities of different hadron species in deep inelastic
scattering off various nuclei. Here the photon-energies are of the
order of 10--20 GeV, with rather moderate $Q^2 \approx 2$ GeV$^2$.
A similar experiment at somewhat smaller energies is currently
performed by the CLAS collaboration at Jefferson Lab
\cite{Brooks}. After the planned energy upgrade they will reach
photon energies of the order of 2--9 GeV and similar virtualities
as in the HERMES experiment.

The interpretation of the observed multiplicity spectra reach from
a possible rescaling of the quark fragmentation function at finite
nuclear density combined with hadron absorption in the nuclear
environment \cite{Accardi} to purely partonic energy loss through
induced gluon radiation of the struck quark propagating thorugh
the nucleus \cite{ArleoWang}. Note that a detailed understanding
of energy loss and hadron attenuation in cold nuclear matter can
provide a useful basis for the interpretation of jet quenching
observed in ultra-relativistic heavy ion collisions at RHIC.

In our approach \cite{falterform} we assume that the virtual
photon interacts with the bound nucleon either directly or via one
of its hadronic fluctuations. The hadronic components of the
photon are thereby shadowed according to the method developed in
Ref. \cite{falterinc} which allows for a clean-cut separation of
the coherent initial state interactions of the photon and the
incoherent final state interactions of the reaction products. The
photon nucleon interaction leads to the excitation of one or more
strings which fragment into color-neutral prehadrons due to the
creation of quark antiquark pairs from the vacuum. As discussed in
a recent work by Kopeliovich et al. \cite{Kopeliovich} the
production time of these prehadrons is very short. For simplicity
we therefore set the production time to zero in our numerical
realization. These prehadrons are then propagated using our
semi-classical coupled-channel transport theory.

After a formation time, which we assume to be a constant $\tau_f$
in the restframe of the hadron, the hadronic wave function has
build up and the reaction products behave like usual hadrons.
During the formation time we distinguish between the so-called
leading prehadrons which contain (anti-)quarks from the struck
nucleon or the hadronic component of the photon and the
non-leading hadrons that solely contain quarks and antiquarks
created from the vacuum. The leading prehadrons can be considered
as the debris of the initial hadronic state and each interacts
with an effective cross section $\sigma_\mathrm{lead}$ during the
formation time. The sum of the leading hadron cross sections
should resemble approximately that of the initial hadronic state.
The nonleading prehadrons that are newly-created from the vacuum
are assumed to be noninteracting during $\tau_f$. Each time when a
new hadron has formed the total effective cross section of the
final state then rises like in the approach of Ref. \cite{Ciofi}.

\begin{figure}[h]
\vspace*{-0.0cm}
 \centering{\includegraphics[width=10.0cm]{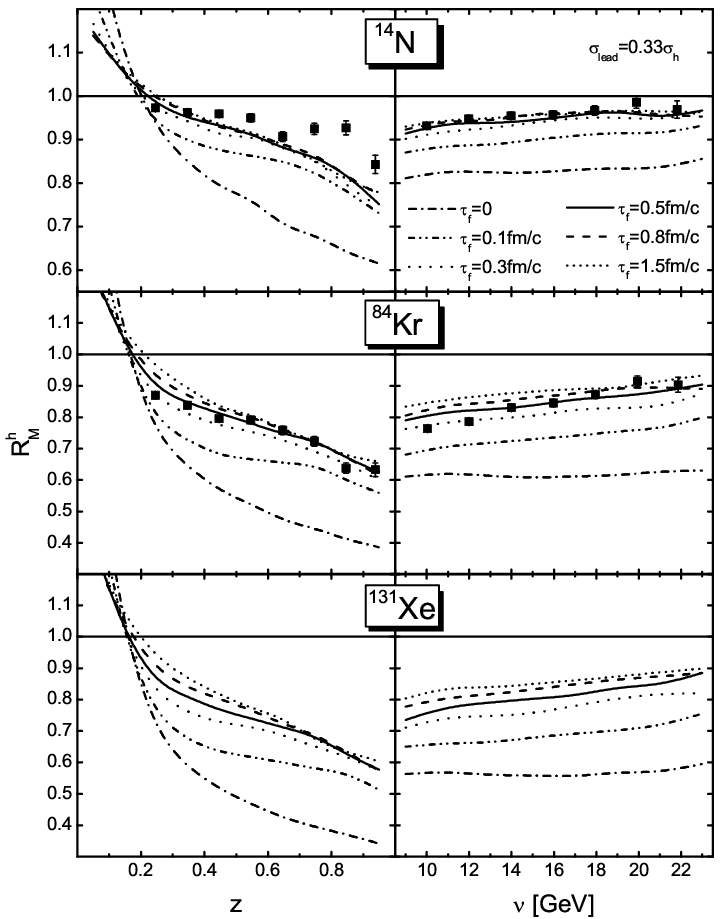}}

\begin{minipage}[t]{12 cm}
\caption{HERMES multiplicity of produced hadrons normalized to the
deuterium as a function of photon-energy $\nu$ (right) and of the
energy of the produced hadrons relative to the photon-energy, $ z
= E_h/\nu$. The curves are calculated for different formation
times given in the figure (from \cite{falterproc}).}
\label{formtime}
\end{minipage}
\end{figure}

Due to our coupled channel-treatment of the final state
interactions the (pre)hadrons might not only be absorbed in the
nuclear medium but can produce new particles in an interaction,
thereby shifting strength from the high to the low energy part of
the hadron spectrum. Our calculated multiplicity ratios of charged
hadrons are shown in Fig. \ref{formtime} in comparison with the
experimental HERMES data \cite{Hermes}. The figure clearly shows
that the observed hadron multiplicities can be described only with
formation times $\tau_f> \approx 0.3$ fm. The curves obtained with
larger formation times all lie very close together. This is a
consequence of the finite size of the target nucleus: if the
formation time is larger than the time needed for the preformed
hadron to transverse the nucleus, then the sensitivity to the
formation time is lost. In \cite{falterform} we have also shown
that the $z$-dependence on the left side exhibits some sensitivity
to the interactions of the leading hadrons during the formation;
the curves show in Fig.\ \ref{formtime} are obtained with a down
scaled leading hadron cross section of $\sigma_\mathrm{lead}=0.33
\sigma_h$, where $\sigma_h$ is the 'normal' hadronic interaction
cross section.

\section{Conclusions}\label{concl}
In this lecture note we have illustrated with a few examples that
photonuclear reactions can yield information that is important and
relevant for an understanding of high density -- high temperature
phenomena in ultrarelativistic heavy-ion collisions. Special
emphasis was put in this article not so much on the theoretical
calculations of hadronic in-medium properties under simplified
conditions, but more on the final, observable effects of any such
properties. We have discussed both reactions with hadronic final
states as well as those with FSI-free lepton states. While in the
former reactions the FSI on the final decay products usually is
strong enough to mask the in-medium spectral properties of the
hadron under study, in the latter case (semi-hadronic or leptonic
final states) the situation is much more encouraging. We have, for
example, shown that in photonuclear reactions in the 1 - 2 GeV
range the expected sensitivity of dilepton spectra to changes of
the $\rho$- and $\omega$ meson properties in medium is as large as
that in ultrarelativistic heavy-ion collisions. We have also
illustrated that the analysis of hadron production spectra in
high-energy electroproduction experiments at HERMES gives
information about the interaction of forming hadrons with the
surrounding hadronic matter. This is important for any analysis
that tries to obtain signals for a QGP by analysing high-energy
jet formation in ultrarelativistic heavy-ion reactions.

\section*{Acknowledgement}
This work has been supported by the Deutsche
Forschungsgemeinschaft, the BMBF and GSI Darmstadt.

\end{document}